\documentstyle[11pt,newpasp,epsf,twoside]{article}
\begin{document}

\title{The Circumstellar Environment of the TTS SU Aurigae}
\author{J.M.~Oliveira \altaffilmark{1}, B.H.~Foing}
\affil{Solar System Division, ESA Space Science Department, ESTEC/SO, Postbus~299, 2200~AG~Noordwijk, The Netherlands}
\author{J.Th.~van Loon}
\affil{Institute of Astronomy, Madingley Road, Cambridge CB3~0HA, United Kingdom}

\altaffiltext{1}{Centro de Astrof\'{\i}sica da Universidade do Porto, Rua das Estrelas s/n, 4150~Porto, Portugal}

\begin{abstract}
In this contribution we investigate how a disk determines the physical and geometrical properties of the circumstellar environment of the T Tauri SU Aurigae. Our model of the spectral energy distribution of this star includes the central young star, a flat black disk and a diffuse envelope. We also describe the inner interaction region between the disk and the star, the magnetosphere, by analysing the accretion and wind signatures in several spectral lines observed during the MUSICOS~96 campaign.
\end{abstract}

\section{Introduction}

\indent Young stellar objects have extremely complex circumstellar environments: material in an infalling envelope surrounding the central star falls onto a circumstellar disk, dissipating angular momentum; magnetic fields can then channel the disk material onto the star by way of accretion funnels (magnetospheric accretion), causing hot shock regions on the stellar photosphere.\\
\indent Classical T Tauri Stars (CTTS) are low-mass pre-main sequence stars. These stars are active, with broad energy distributions and Balmer and metallic (permitted and forbidden) line emission. CTTS are class II objects, i.e. the systems include a central young star and an accretion disk. The presence of the disk was inferred from the IR and UV excesses in the spectral energy distributions and it can also account for some of the photometric variability and optical veiling; for some stars the disk has also been imaged directly.\\
\indent Magnetospheric accretion can explain the photospheric hot spots and the He~I spectral line, as a shock region at the stellar surface results in the release of kinetic energy. Spectroscopic evidence of magnetospheric accretion are the redshifted absorption components on several spectral lines. The spectra of T Tauri stars also present blueshifted absorption features that are signatures of stellar or disk winds.\\
\indent SU Aurigae is either classified as a CTTS (e.g. Bouvier et al. 1993) or as the prototype of a separate class of young stellar objects (Herbig \& Bell 1988). The star's spectral type is G2 (Herbig 1960), earlier than most CTTS, and it is an exceptionally fast-rotator, v $\sin i \sim 66$~km~s$^{-1}$ (Hartmann \& Stauffer 1989). Its photometric period has been rather difficult to determine, but it is usually considered to be close to 3~days. From our data set, we favour a somewhat shorter period of 2.5-2.8~days (Oliveira et al. 2000).\\
\indent In this contribution we describe the SU~Aur system, from two different viewpoints: i) using the spectral energy distribution to characterize the larger scale circumstellar environment and ii) probing the disk-star interaction by analysing the inner magnetosphere, using results from our multi-site spectroscopic campaign on this target.

\section{The Outer Circumstellar Environment: Disk \& Envelope}

\begin{figure}[ht]
\label{f1}
\plotfiddle{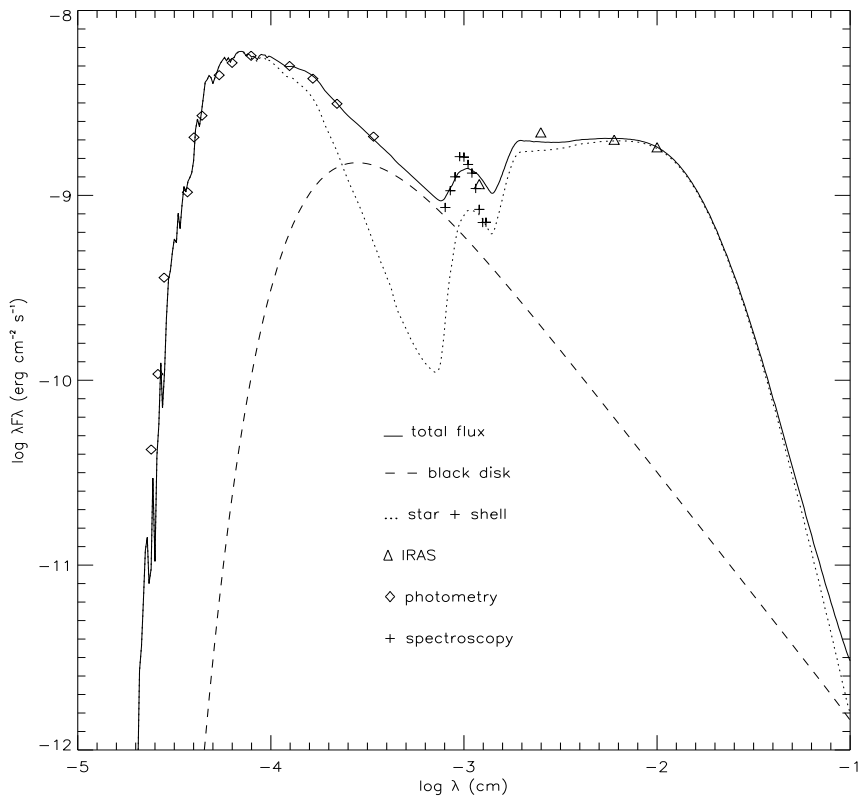}{10cm}{0}{60}{55}{-180}{0}
\caption{SED for SU AUR: $\diamond$ represent photometric (Bertout et al. 1989), + spectroscopic (Hanner et al. 1998) and $\triangle$\ IRAS measurements. The resulting model is plotted (full line) as well as the star~+~shell (dotted line), computed with the MoDust code, and the flat black disk (dashed line) contributions. The fit obtained is quite good, with exception of the 12-20~$\micron$ range (silicate features and detailed treatment of chemistry is required) and the UV excess (shock region emission).}
\label{f1}
\end{figure}

Accretion disks can explain two of the most important peculiarities of the Spectral Energy Distribution (SED) of the CTTS: the UV and IR excesses. The dust grains in the disk are heated either radiatively by the central star or by accretion heating (via the loss of gravitational energy as the gas spirals through the disk to the star) and they emit in the IR. As the visual extinction is relatively low for most CTTS the dust distribution has to be flattened, as it was proved recently by HST images of several CTTS. The UV excess can originate at the shock regions that form where the material accretes from the disk onto the stellar surface, causing the release of large amounts of kinetic energy.\\ 
\indent The SEDs of some CTTS seem to display more far-IR excess than expected from a disk. Several methods have been used to model these so called flat spectrum: adding to the system an accreting envelope component (Calvet et al. 1994) or considering a flared disk, i.e. a disk whose vertical height varies with the radial distance (Kenyon \& Hartmann 1987).\\
\indent  We review here the SED of SU Aur, using photometry from the literature and redetermining the IRAS fluxes. This SED is relatively flat at the far-IR.  It seems to have two contributions: a cool and a hot dust component. We find that the cool component can be well represented by a spherical dust shell at some distance from the star, whilst a disk closer to the star can account for the hot component. We modelled this disk as a flat, black disk (Beckwith 1999). The fit obtained is quite good (within 0.05 in units of $log \lambda F_{\lambda}$), with the exception of the silicate features at 10 and 18~$\mu$m, that demand a deeper study into the dust properties in the disk and envelope. Our model also does not fit properly the UV excess: the emission at those wavelengths is believed to be related with the accretion energy released at the footpoints of the accretion funnels at the stellar surface. The resulting fit to the SED is shown in Fig.~1 and the main model parameters are given in Table~1. 

\begin{table}
\vspace*{-0.5cm}
\begin{center}
\caption{\centering The model parameters for the SED of SU~Aur}
\vspace*{0.5cm}
\begin{tabular}{|c|c|c|c|c|c|}
\hline
\multicolumn{2}{|c|}{star} & \multicolumn{2}{c|}{shell} & \multicolumn{2}{c|}{disk}\\
\hline
distance & 150~pc & R$_{in}$ & 200~R$_{*}$ & R$_{in}$ & 3~R$_{*}$\\ 
L$_{*}$ & 9.5~L$_{\odot}$ & R$_{out}$ & 10$^{6}$~R$_{*}$ & R$_{out}$ & 3~$\times$~10$^{4}$~R$_{*}$\\
R$_{*}$ & 3.4~R$_{\odot}$ & $\rho_{in}$ & 5~$\times$~10$^{-17}$ & inclination & 60$^{\circ}$\\
T$_{eff}$ & 5500~K & &(g cm$^{-3}$) & & \\
\hline
\end{tabular}
\end{center}
\label{t1}
\end{table}

\section{The Inner Circumstellar Environment: Magnetosphere \& Spectral Line Variability}

Shu et al. (1994) proposed a magnetospheric model where the accretion disk is disrupted at a few stellar radii from the star by a stellar dipole magnetic field. At the truncation point, the ionized disk material is loaded either onto inner closed magnetic field lines, accreting thereby onto the stellar photosphere, or it is loaded onto outer open magnetic field lines that can drive a disk-wind flow. Johns \& Basri (1995) suggested that in the case of SU Aur the rotational and magnetic axes are mis-aligned, causing the observed periodicity ($\sim$~3~days) in the wind and accretion flows and their anti-phase behaviour.\\ 
\indent We used the data set obtained in the MUSICOS~96 campaign to characterize the variability time scales of several lines. In Fig.~2 we show the average and variance profiles of H$\alpha$, H$\beta$, Na~I~D and He~I~D3. Using the cross-correlation function, we searched for time lags between the variability of these spectral lines. We have found that the variations in the red wing of H$\beta$ are out of phase with the variations in the blue wing of H$\beta$, by slightly more than half of the rotational period (Fig.~3~(a)). Thus as expected in an oblique magnetosphere the accretion funnel and the wind components are out of phase.\\

\begin{figure}[ht]
\plotfiddle{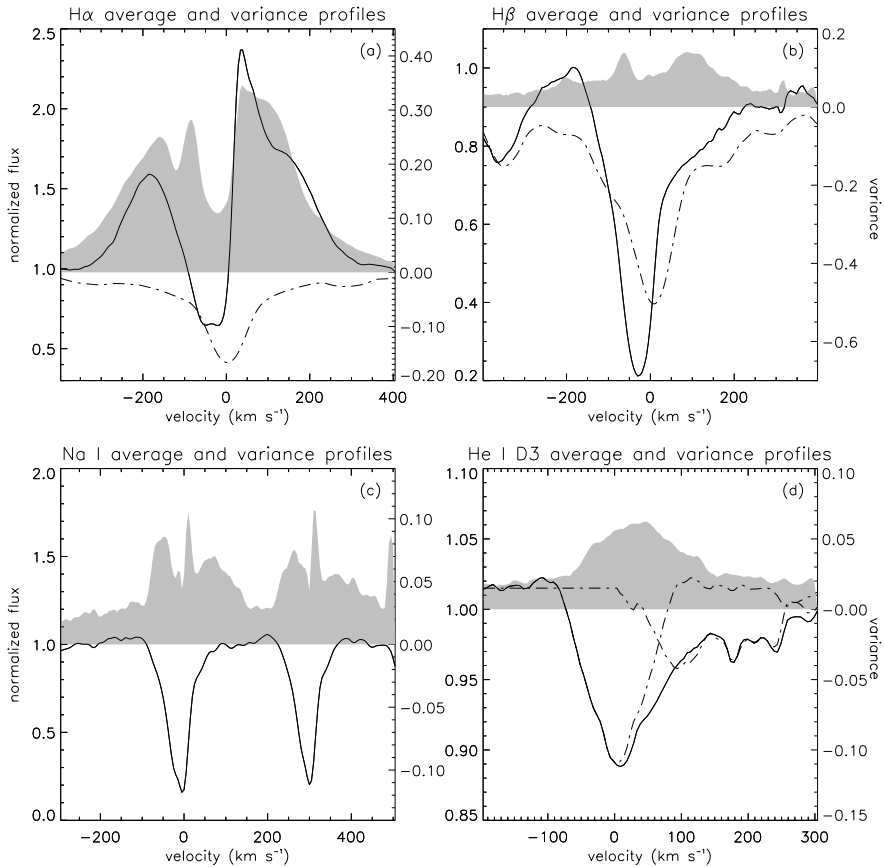}{11cm}{0}{78}{78}{-170}{0}
\caption{The average (line plots) and variance (shaded plots) of H$\alpha$\ (a), H$\beta$\ (b), Na~I~D (c) and He~I~D (d). The y-axis range on the left side is for the average and on the right side for the variance.}
\end{figure}

\indent The He~I~D3 line is in absorption in SU Aur, contrary to what is observed for most of the other CTTS. Still, we have also decomposed this profile in two components, one centered at rest velocity and another centered at about 80~km s$^{-1}$. We have found that there might be a small time lag between these components, suggesting that the first component might originate at the surface and the redshifted component in the accretion column. We have also found that the variability in the He~I~D3 and the red wing of Na~I~D are almost in synchrony (Fig.~3~(c,d)) and it propagates in a few hours to the absorption components in the red wings of H$\beta$\ and H$\alpha$ (Fig.~3~(b)).\\
\indent We thus observe variability that propagates from the inner magnetosphere outwards. Such behaviour cannot be caused by non-steady accretion nor by simple occultation effects in the oblique magnetospheric model. We propose that this can only occur if the magnetic field lines are azimuthally distorted due to differential rotation between the star and the disk (Oliveira et al. 2000). In this situation, the time symmetry for ingress/egress of occultation events is broken and then the variability of the higher energy lines precedes that of the lower energy lines. 

\begin{figure}[h]
\plotfiddle{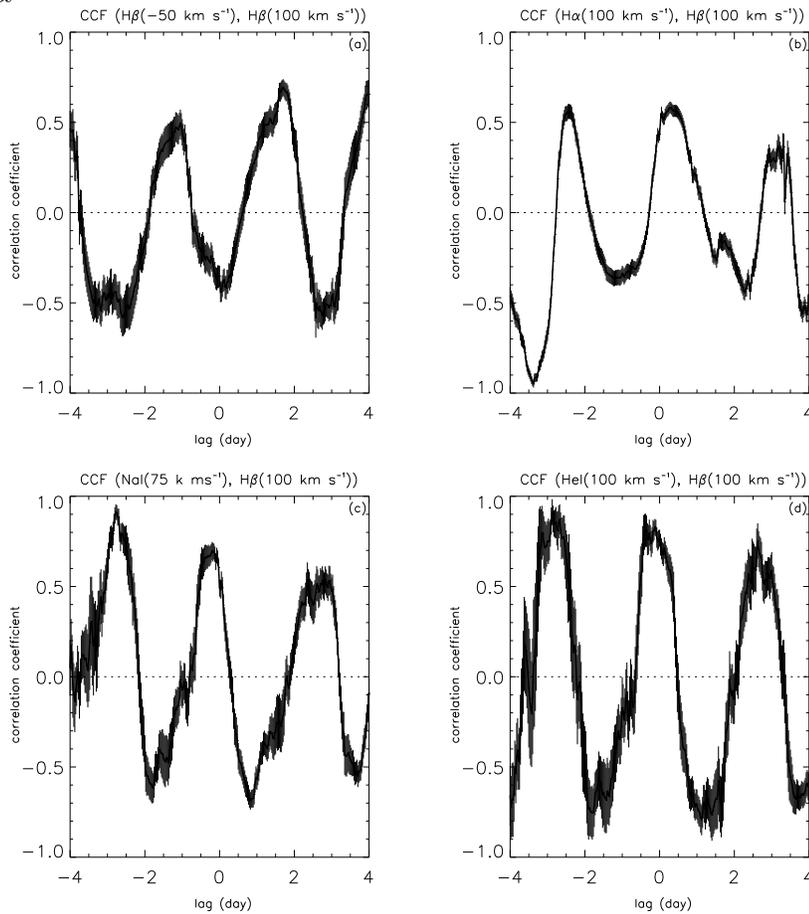}{11.5cm}{0}{65}{65}{-180}{0}
\caption{Cross-Correlation as a function of time lag. The quasi-periodic shape of the CCF reflects the periodicity of the spectral lines. The variability in He~I~D3 and Na~I~D slightly precedes the variability in H$\beta$; the variability in H$\beta$ precedes the variability in H$\alpha$. Thus, the observed variability seems to propagate from the inner magnetosphere to the outer regions.}
\end{figure}

\section{Conclusions}

\indent The enhanced activity observed in SU Aur can be explained by the presence of an accretion disk and spherical dust shell. We describe the SED of SU Aur as including the star, a flat black disk (a hot dust component to account for the near-IR excess emission) and a diffuse envelope (a cold dust component to account for the far-IR excess). The model is rather simple but it gives a very good agreement with the observations available.\\
\indent Probing the inner magnetosphere, we observe a time-lagged behaviour of the variability of spectral lines with different excitation conditions that is not compatible with the standard magnetospheric models, even when the magnetic dipole is inclined with relation to the rotation axis. The behaviour we observe requires that the magnetic field lines are azimuthally distorted probably due to differential rotation in the star~+~disk system. Obviously, more stars need to be observed quasi-continuously to allow a generalization of the results presented here.

\acknowledgements
JMO acknowledges the support of the {\it Funda\c{c}\~{a}o para a Ci\^{e}ncia e Tecnologia} (Portugal) under the grant BD9577/96. We acknowledge Alex de Koter from the Astronomical Institute, Amsterdam for the use of his MoDust code.

\end{document}